# Strong and engineerable optical anisotropy in easily integrable epitaxial SrO(SrTiO$_3$)$_N$ Ruddlesden–Popper thin layers

M.O. Bounab[1], C. Furgeaud[1], S. Cueff[1], L. Berguiga[1], R. Bachelet[1], M. Bouras[1,3], L. Pedesseau[4], J. Even[4], L. Largeau[2], G. Saint-Girons[1]*

[1]INL-UMR5270/CNRS, 36 avenue Guy de Collongue, 69134 Ecully cedex (France)

[2] Centre de Nanosciences et de Nanotechnologies, Universite Paris-Saclay, CNRS, Palaiseau (France)

[3] present affilliation: ISOM - Dept. Ing. Electronica, ETSIT, Univ. Politecnica, 28040 Madrid, (Spain)

[4] Univ Rennes, INSA Rennes, CNRS, Institut FOTON – UMR 6082, F-35000 Rennes, France

*Corresponding author email: guillaume.saint-girons@ec-lyon.fr

## Abstract

Optical anisotropy is a key property for various photonic devices. While bulk anisotropic materials used for photonic applications are relatively scarce, and often tricky to synthesize as thin films, the optical losses as well as complex structuration of anisotropic metamaterials limit their integrability in photonic devices. On the basis of ellipsometric measurements coupled with reflectance, we show here that Ruddlesden-Popper SrO(SrTiO$_3$)$_N$ phases (STO-RP$_N$), epitaxial thin films consisting of a SrTiO$_3$ lattice in which one SrO atomic plane is inserted every N unit cells, exhibit strong dichroism and birefringence over a wide spectral range. More, this anisotropy is tuneable by modifying the RP order N. Unlike most other anisotropic materials described in the literature, STO-RP$_N$ thin layers are fabricated using industry-standard growth processes. As they can be epitaxially grown on Si and GaAs using SrTiO$_3$ templates, our work paves the way for their compact integration on these photonic platforms.

## 1. Introduction:

Optical anisotropy is at the heart of many essential optical functions. Materials with high linear dichroism, whose absorption strongly depends on light polarization, are used to design polarization sensitive detectors,[1,2,3,4] which enable many applications ranging from polarization spectroscopy imaging to sensing, guiding and communication systems. [1,2,3,4,5,6,7] They have for instance been used as saturable absorber in ultrafast polarized pulsed laser sources.[8] They have also been used in synaptic devices for neuromorphic applications[9] and in information encryption devices.[10] Birefringent materials, having anisotropic refractive index, enable light polarization control and are therefore used in fundamental optical devices based on phase retardation, such as waveplates, phase-matching elements, modulators or polarization-controlled light emitting devices.[11,12,13]. They also enable the generation of low-loss optical modes[14] which have led to the demonstration of ultraconfined light coupling,[15] resonators with high quality factors[16] and bound states in the continuum.[17] Such exotic optical modes present significant opportunities for optical engineering. For instance, anisotropic materials support polaritons with hyperbolic dispersion, enabling high localization of electromagnetic energy and presenting a high density of photon states, useful for infrared imaging and detection, thermal energy transfer, molecule detection and more.[15] Optically anisotropic materials thus offer unique opportunity to design novel functionalities for photonic integrated circuits, with potential applications ranging from sensors, displays and medicine to non-linear and quantum optics. [12,18,19]. In the latter domain, the control of photon polarization enabled by anisotropic materials is a key function for proper operation of quantum gates, the building blocks of quantum computing systems.[20] Anisotropic materials also offer new opportunities in the field of quantum plasmonics. Thus, for example, hBN anisotropy leads to an enhancement of the local density of states in plasmonic tunnel junctions, enabling an increase in their emission efficiency.[19]

The fabrication of photonic devices based on these functions requires optically anisotropic materials to be integrated compactly, in the form of thin films, into photonic structures. In addition, integrability with silicon based photonic circuits, the mainstream technology for future developments in the photonic field,[21] is advisable. In the end, flexibility in controlling the optical anisotropy is required for device design. Regarding this last point, so-called optical metamaterials based on top-down or multilayer micro/nanostructuration are promising: metamaterial structuration provides significant flexibility for permittivity engineering, and high linear dichroism has been generated from nanowires or nanorods,[22,23] as well as from structured metasurfaces.[24] Similarly, strong birefringence has been achieved using arrays of optical antennas,[25] multi-slotted Si waveguides,[26] arrays of semiconductor nanowires[27] or multilayer polymeric thin films.[12] However, the metamaterial morphology (surface pattern, surface metal structures, thick superlattices,…) complicates their integration into devices.[28]

Homogeneous materials with intrinsic anisotropy are much easier to integrate into photonic devices, and offer larger flexibility for optical design.[29] Yet quite few solid crystals exhibit significant optically anisotropy. Strong dichroism has been reported for several 2D materials (typically fabricated by exfoliation from bulk) such as black phosphorus[2] and black AsP[30], ReS$_2$,[1] ReSe$_2$,[5] GeSe[3], GeAs,[31] GeAs$_2$,[4] GeSe$_2$[32] and PdSe$_2$[33], as well as in diphenyl anthracene (DPA)[6] synthetized by chemical routes and in BaTiS$_3$, fabricated using the vapor transport method.[34] Strong birefringence has been reported for calcite[35] (typically synthesized by chemical routes), a few borates and phosphates such as $\alpha$-BaB$_2$O$_4$,[36] Ba$_2$Ca(B$_3$O$_6$)$_2$,[37] and Sn$_2$PO$_4$I[38] synthesized from top seeded solutions or Czochralski growth, iodates based composite materials such as HfH$_2$(IO$_3$)$_2$[39] and Sc(IO$_3$)$_2$(NO$_3$)[40] elaborated by hydrothermal synthesis, BaTiS$_3$[30] and Sr$_{9/8}$TiS$_3$[41] sulphides, synthesized by the vapor transport method, some halide perovskites[42] such as CsPbBr$_3$,[43] fabricated by chemical routes, and several van der Waals crystals[44,45] (MoS$_2$,[46] h-BN,[45,47] Ta$_2$NiS$_5$,[48] $\alpha$-MoO$_3$,[49] GaSe[50]) typically fabricated from powder or by exfoliation. However, in these materials, the optical anisotropy is a bulk property and cannot be engineered to meet device requirements. Additionally, most of these materials cannot be grown as high quality thin films and some of them are chemically unstable. The growth techniques used to synthesize them are not, or only to a limited extent, photonics industry-standard processes, which limits their integrability in commercial photonics systems. A homogeneous material exhibiting strong and tunable optical anisotropy, fabricated using an industry-standard process, is therefore still lacking.

Perovskite oxides of general formula ABO$_3$ can host a variety of cations with minimal structural changes.[51] This chemical flexibility provides access to a wide range of properties, including ferroelectricity, magnetism, conductivity, dielectricity, etc. Furthermore, many of these materials remain stable against periodic insertion of additional AO planes every N unit cells along the <001> directions,[52] leading to the formation of so-called N$^{th}$ order Ruddlesden-Popper (RP) phases of general formula A$_{N+1}$B$_N$O$_{3N+1}$.[53] Hence, the N$^{th}$ order Ruddlesden-Popper phase based on SrTiO$_3$ (STO) (formula Sr$_{N+1}$Ti$_N$O$_{3N+1}$), referred to as STO-RP$_N$ in the following, is formed by inserting an additional SrO atomic plane every N STO unit cells along the [001] direction. This structuration at the atomic plane scale is an interesting lever to engineer the material properties. Among others, STO-RP$_N$ phases present promising thermoelectric properties[54] and outstandingly low dielectric losses.[55] STO-RP$_N$ phases with N > 3 are unstable against demixion,[56] but molecular beam epitaxy (MBE), allowing the implementation of highly non-equilibrium layer-by-layer growth strategies, has enabled the fabrication of excellent quality STO-RP$_N$ thin films on STO substrates, with N= 1 and up to 10.[57,58] Literature on the optical properties of RP phases is very sparse. In Ref.59, the authors measure the bandgaps of STO-RP$_N$ thin layers (N = 1,…5 and N=10) grown by oxide MBE substrates using ellipsometry, transmission and cathodoluminescence assuming that RP phases present an isotropic optical response. Ref.60 emphasizes the anisotropy of the electronic band structure of STO-RP$_1$ thin layers. Finally, bandgap measurements carried out on polycrystalline STO-RP$_1$ and STO-RP$_2$ powder samples fabricated by standard ceramic routes are reported in Ref.61.

As the crystal structure of STO-RP$_N$ phases is anisotropic (space group I4/mmm), one might expect their optical response to be as well, as predicted by several theoretical works.[56,62,63,64] Leveraging a method combining normal incidence reflectivity and ellipsometry, we present here the first experimental evidence that STO-RP$_N$ layers (N = 1-5), grown by MBE, a process used in industry, are indeed uniaxial birefringent materials with an optical axis perpendicular to the (001) surface. Their birefringence $|\Delta n|$ ( defined as $|n_\parallel - n_\perp|$, where $n_\parallel$ and $n_\perp$ respectively designates the real part of the in- and out-of-plane components of the complex refractive index) and dichroic ratio $k_\perp / k_\parallel$ (where $k_\parallel$ and $k_\perp$ respectively designates the imaginary part of the in- and out-of-plane components of the complex refractive index) are comparable to the largest value reported in the literature.

## 2. Results and discussion

A summary of the structural characterizations carried out on the STO-RP$_N$ samples (N = 1-5) by High Angle Annular Dark Field Scanning Transmission Electron Microscopy (HAADF-STEM), X-ray diffraction (XRD) and Reflection High Energy Electron Diffraction (RHEED) is shown in Fig. 1.

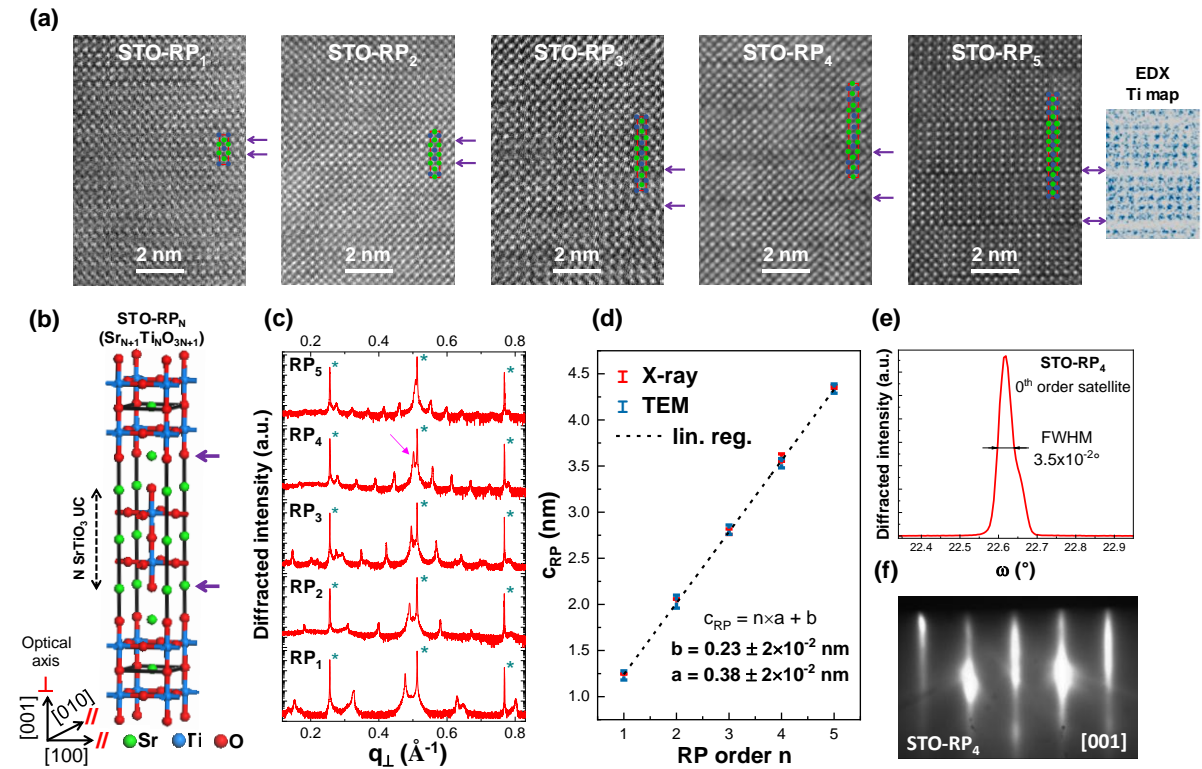

*Figure 1: Structural characterizations. (a) HAADF-STEM micrographs of the samples. (b) STO-RP$_N$ lattice (N = 2 in the example): black lines delimit the unit cell, and purple arrows indicate the extra SrO planes every n SrTiO$_3$ unit cells. The ordinary axes (resp. extraordinary axis) are indicated by // (resp. ⊥). (c) XRD radial scans recorded on the samples (q$_\perp$ designates the (00l) reciprocal space vector, the blue stars indicate substrate reflexions and the pink arrow indicate the 0$^{th}$ order satellite of the STO-RP$_4$ layer). (d) Lattice parameters (c$_{RP}$) extracted from X-ray scans (red error bars) and from STEM images (blue error bars) as a function of the RP order N. (e) XRD ω-scan recorded around the 0$^{th}$ order satellite of the STO-RP$_4$ layer (indicated in pink in (c)). (f) RHEED pattern recorded along the [100] azimuth of the sample surface at the end of the growth of the STO-RP$_4$ sample.*

These characterizations attest for the good RP ordering in the samples: the XRD pattern show that the layers are epitaxial, (001)-oriented, of good crystalline quality, and free of parasitic phases. Some vertical SrO-SrO stacking faults accommodate the slight stoichiometry imbalances occurring during the growth due to effusion cell instability, as commonly observed in epitaxial RP phases (Fig. S5).[57,58,65] For all samples, the XRD ω-scans (Fig.1(c) for the STO-RP$_4$ sample, Supporting Information (Fig. S4) for the others) show low mosaicity of the order of a few 10$^{-2}$.° The out-of-plane lattice parameters of the five STO-RP$_N$ samples extracted from the XRD measurements, namely c(STO-RP$_1$)=12.5±0.15 Å, c(STO-RP$_2$)=20.4±0.13 Å, c(STO-RP$_3$)=28.1±0.15 Å, c(STO-RP$_4$)=35.6±0.1 Å and c(STO-RP$_5$)= 43.5±0.9 Å, are consistent with that reported in literature,[66] and with that extracted from the TEM images of Fig.1(a), as shown in Fig.1(d). According the RHEED patterns displayed in Fig.1(f) and in the Supporting Information (Fig.S1), the samples have smooth, two-dimensional crystalline surfaces.

As mentioned above, STO-RP$_N$ crystallographic anisotropy and theoretical works available in literature[56,62,63,64] suggest these materials are uniaxial birefringent, with an optical axis perpendicular to the (001) surface. However, this optical anisotropy has yet to be confirmed experimentally. For a uniaxial anisotropic material, the component of the refractive index in the (001) plane, designated as $\tilde{n}_\parallel = n_\parallel + ik_\parallel$, differs from its component along the [001] optical axis, designated as $\tilde{n}_\perp = n_\perp + ik_\perp$. Ellipsometry, and in particular generalized ellipsometry, is a suitable technique for characterizing anisotropic materials,[67] except in the configuration expected for the STO-RP$_N$ samples. Indeed, in such a configuration and despite the material anisotropy, a good fit of the ellipsometric measurements can be achieved using an isotropic refractive index $\tilde{n}_{eff} = n_{eff} + ik_{eff}$, with $\tilde{n}_{eff}$ dependent on $\tilde{n}_\parallel$ et $\tilde{n}_\perp$,[68,69] as illustrated in the Supporting Information (section SIII). In order to confirm STO-RP$_N$ optical anisotropy, we therefore implemented an analysis combining variable angle spectroscopic ellipsometry (VASE) and reflectivity measurements carried out at zero incidence angle. It is based on the fact that the zero incidence reflectivity of an uniaxial birefringent material having its optical axis perpendicular to its surface is equal to that of a material of refractive index $\tilde{n}_\parallel$ (in plane component refractive index), since the light polarization at normal incidence is always perpendicular to the optical axis.[70] If such a material is deposited in the form of a thin layer of thickness $d$ on an isotropic substrate of refractive index $\tilde{n}_s$, the zero incidence reflectivity thus reads[71]:

$$R_{0°}(\tilde{n}_\parallel) = \left| \frac{\dfrac{\tilde{n}_\parallel - 1}{n_0 + 1} + \dfrac{\tilde{n}_s - \tilde{n}_\parallel}{\tilde{n}_s + n_0} e^{-2i\tilde{n}_\parallel \frac{2\pi d}{\lambda}}}{1 + \dfrac{\tilde{n}_\parallel - 1}{\tilde{n}_\parallel + 1} \times \dfrac{\tilde{n}_s - \tilde{n}_\parallel}{\tilde{n}_s + \tilde{n}_\parallel} e^{-2i\tilde{n}_\parallel \frac{2\pi d}{\lambda}}} \right|^2 , \quad (Eq. 1)$$

with λ the wavelength.

For an anisotropic material, the refractive index extracted from the fit of the VASE measurements using an isotropic model ($\tilde{n}_{eff}$) differs from $\tilde{n}_\parallel$, so that the zero incidence reflectivity calculated from $Eq.1$ using $\tilde{n}_{eff}$ ($R_{0°}^{calc} = R_{0°}(\tilde{n}_{eff})$) differs from the experimental zero incidence reflectivity $R_{0°}^{exp}$. In contrast, for an isotropic material, the refractive index extracted from the fit of the VASE measurements using an isotropic model corresponds physically to the refractive index of the material, and the zero-incidence reflectivity calculated from $Eq.1$ using this refractive index corresponds to the experimental reflectivity. We performed VASE and zero-incidence reflectivity measurements on the five STO-RP$_N$ samples and on an optically isotropic STO substrate, used as a reference (see Supporting Information (section SIII) for experimental details). For each sample, the VASE measurements were then fitted using isotropic models, and the resulting refractive index was injected into $Eq.1$ (with $d \rightarrow \infty$ for the STO substrate) to determine the calculated zero-incidence reflectivity $R_{0°}^{calc}$. The latter is compared to the experimental zero-incidence reflectivity $R_{0°}^{exp}$ in Fig.2.

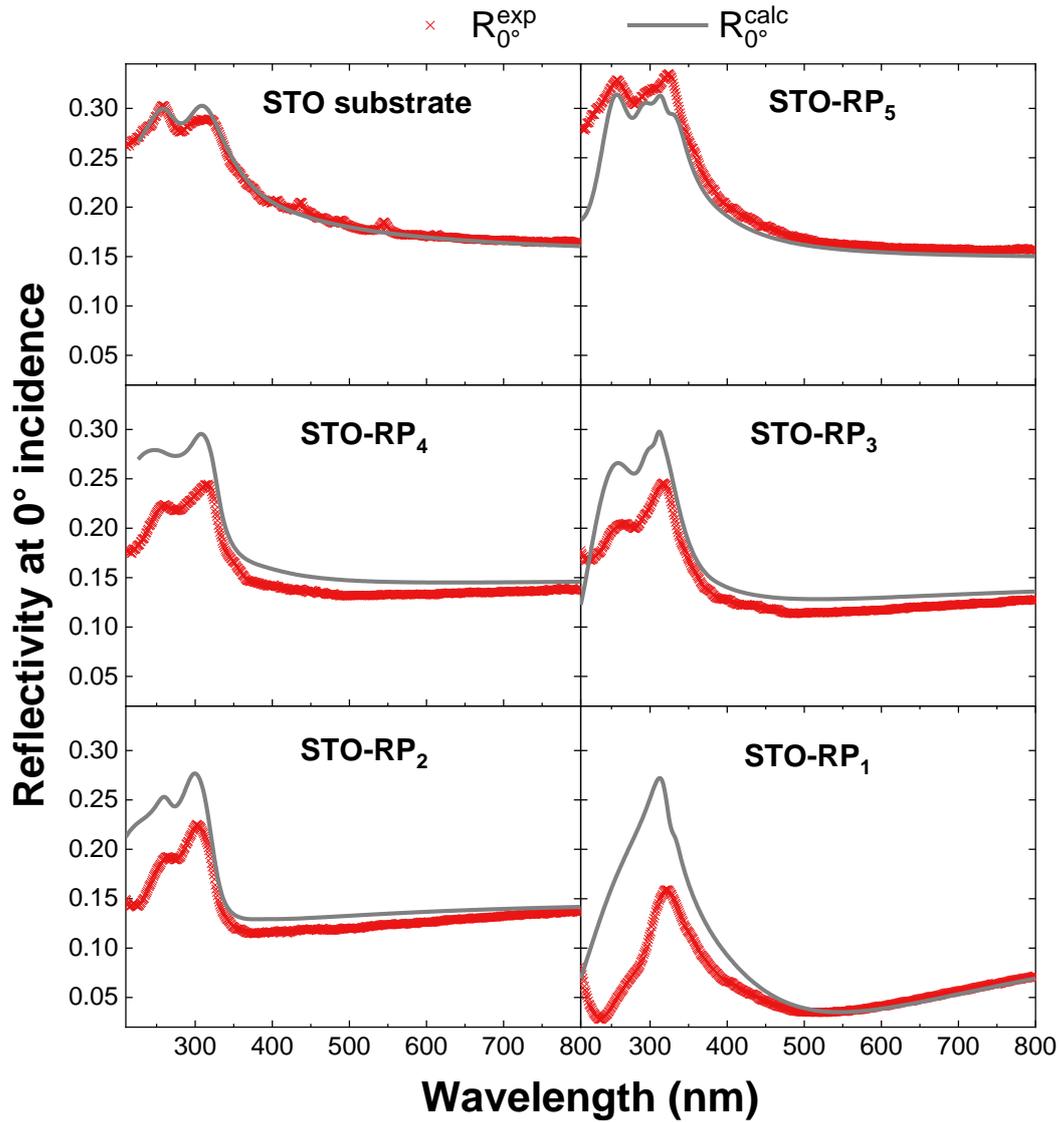

*Figure 2: Experimental demonstration of optical anisotropy.* $R_{0°}^{exp}$ *(red crosses) and* $R_{0°}^{calc}$ *(grey lines) for the five STO-RP$_N$ samples and the reference STO substrate.*

As expected, $R_{0°}^{calc}$ and $R_{0°}^{exp}$ are equal within experimental uncertainty for the isotropic STO substrate. Conversely, $R_{0°}^{eff} \neq R_{0°}^{calc}$ for the five STO-RP$_N$ samples, thereby demonstrating their optical anisotropy. It should be noted that the approach described above does not allow for the exclusion of biaxial anisotropy (two different refractive indices in the two directions of the plane orthogonal to the sample surface). However all the crystallographic data collected for the epitaxially grown thin films are compatible with the reported bulk tetragonal space group and make this hypothesis very unlikely.

More, ellipsometry measurements performed for different angular positions around the axis perpendicular to the sample surface give identical results (not shown here), demonstrating that the optical anisotropy of the STO-RP$_N$ thin films is indeed uniaxial.

To determine the refractive indices of these five samples, VASE and zero-incidence reflectivity measurements were fitted simultaneously with the DeltaPsi2 software (Horiba Scientific), as detailed in the Supporting Information (section SIII). Layer thicknesses were fixed equal to the values measured by XRR, namely 46.2±0.2 nm, 48.4±0.7 nm, 49.3±0.2 nm, 53.9±0.9 nm and 48.7±0.1 nm for samples STO-RP$_5$, STO-RP$_4$, STO-RP$_3$, STO-RP$_2$ et STO-RP$_1$, respectively. The results are shown in Fig.3.

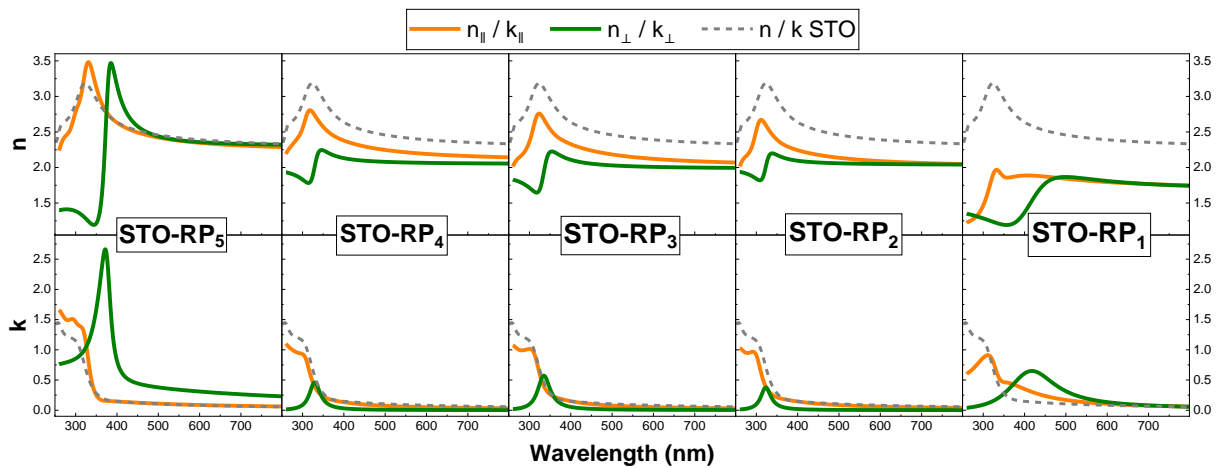

*Figure 3: Refractive indices and extinction coefficients for the five STO-RP$_N$ samples, compared to that of STO.*

For STO, the ellipsometry data have been fitted with a Tauc-Lorentz oscillator-based model similar to that reported in Ref.72, as expected for this semiconductor material (Supporting Information, section SIII). We extract a bandgap energy of 3.22 eV and a refractive index in line with those widely documented in literature.[72] For the STO-RP$_N$ samples, the best fit was obtained by using a model of the same type as that used for STO (corresponding to a semiconducting behavior) for the ordinary component of the refractive index, and a model with one or two Lorentz oscillators (corresponding to a dielectric/insulating behavior) for its extraordinary component (Supporting Information, section SIII). The bandgap energy values extracted from the Tauc-Lorentz model for the ordinary component are 3.33, 3.37, 3.39, 3.42 and 3.60 eV for samples STO-RP$_5$, STO-RP$_4$, STO-RP$_3$, STO-RP$_2$ and STO-RP$_1$ respectively (comparable to those reported in literature[59]): the bandgap energy increases with decreasing RP order. Furthermore, the ordinary and extraordinary components of the ordinary refractive index tend to decrease overall as N decreases.

The theoretical work available in the literature on the band structure of STO-RP$_N$ phases helps to interpret these experimental results. Firstly, they show that for STO, as for RP phases, the bandgap and all electronic transitions with energies below 12 eV involve valence-band O$2p$ and conduction-band

Ti$3d$ states.[56,62,63,64,73,74,75] Sr orbitals have no direct influence on these transitions. However, the intercalation of SrO layers in the perovskite structure induces TiO$_2$ chain rupture in the RP phases along the [001] axis.[59,64] This rupture is at the origin of the anisotropy of the dielectric response of the RP phases: in the (001) plane, the Ti-O chains are not broken and the RP phases therefore have, like STO, a large-bandgap semiconductor behavior, whose refractive index can be described with a Tauc Lorentz-type model. In contrast, along the [001] axis, the Ti-O chains are broken. This rupture reduces hybridization between these O$2p$ and Ti$3d$ orbitals.[63] It therefore prevents electrical conduction along the [001] axis, and we attribute to it the dielectric character of the RP phase response in the extraordinary direction. It also reduces the spatial extension of electronic states involving the O$2p$ and Ti$3d$ orbitals.[59,62,63] This localization of electronic states between the additional SrO planes leads to an increase of the electronic confinement as the order of the RP N decreases, causing a bandgap energy increase, as shown theoretically[59,62,63] and experimentally[59] in literature and confirmed by our measurements. Finally, the reduced hybridization of the O$2p$ and Ti$3d$ orbitals reduces the oscillator strength of the associated transitions and consequently reduces the refractive index when N decreases, as predicted theoretically in ref.63 and confirmed experimentally by our measurements.

Our measurements show that STO-RP$_N$ thin films are highly anisotropic in the UV and visible ranges. Their absolute birefringences ($|\Delta n| = |n_\parallel - n_\perp|$) are compared to those of the most anisotropic materials in this spectral range according to literature (namely some van der Waals crystals[45,46,47,48,49] and quasi-1D chalcogenides[34,41]) in Fig.4.

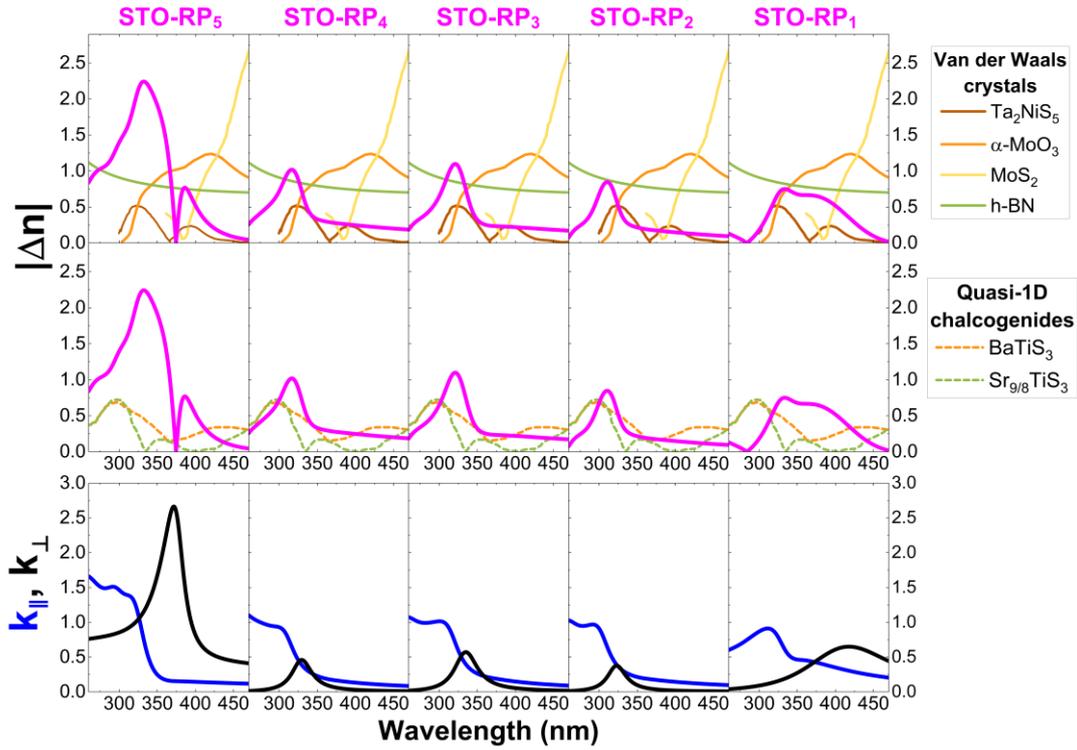

*Figure 4: Birefringence of the STO-RP$_N$ layers compared to that of materials with the highest anisotropy in the UV-visible spectral range.* Top panel: $|\Delta n| = |n_\parallel - n_\perp|$ for the STO-RP$_N$ layers (pink), for the van der Waals crystals Ta$_2$NiS$_5$,[48] α-MoO$_3$,[49] MoS$_2$[46] and h-BN[45,47], and for the quasi-1D chalcogenides BaTiS$_3$[34] and Sr$_{9/8}$TiS$_3$.[41] Bottom panel: In plane and out of plane extinction coefficients for the five STO-RP$_N$ samples.

STO-RP$_5$ birefringence significantly exceeds that of these materials between 260 and 380 nm, with a peak value of 2.25 at 340 nm. This value, the highest reported in the UV range, is close to the highest birefringence reported in the literature at any wavelength, namely $|\Delta n| \approx 3$ at 675 nm for MoS$_2$.[46] STO-RP$_N$ birefringence is close to or above 0.5 from 260 to 450 nm. However, the interest of this high birefringence for integration into photonic structures is mitigated by the significant absorption of STO-RP$_N$ samples, which exhibit extinction coefficients between 0.1 and more than 2 throughout the spectral range of high anisotropy (Fig.4, bottom panel). Even more interestingly, the STO-RP$_N$ also exhibits a very strong linear dichroism. In the literature, the latter is commonly characterized by the dichroic ratio $\frac{k_{max}}{k_{min}}$, where $k_{max}$ (respectively $k_{min}$) is the largest (respectively smallest) component of the extinction coefficient. The dichroic ratios $\frac{k_{max}}{k_{min}}$ of the STO-RP$_N$ samples are compared with those of the materials with the highest linear dichroism according to literature (namely 2,6-diphenyl anthracene (PDA),[6] GeAs,[31] BaTiS$_3$,[34] GeSe[3] and PdSe$_2$[33]) in Fig.5.

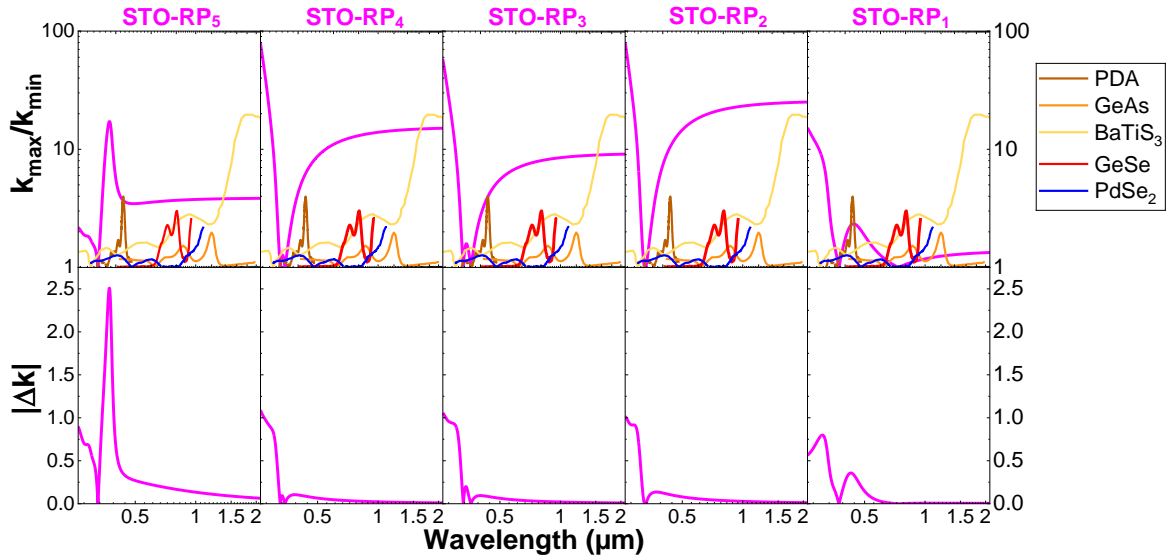

*Figure 5: Dichroic ratio of the STO-RP$_N$ layers compared to that of the materials with the highest dichroic ratios. Top panel: $\frac{k_{max}}{k_{min}}$ for the STO-RP$_N$ layers (pink) and for PDA,[6] GeAs,[33] BaTiS$_3$,[34] GeSe[3] and PdSe$_2$[33] Bottom panel: absolute linear dichroism $|\Delta k| = |k_\parallel - k_\perp|$ for the five STO-RP$_N$ samples.*

The dichroic ratios of the STO-RP$_N$ thin films are the highest reported in the UV range: they exceed 50 around 270 nm for STO-RP$_4$, STO-RP$_3$ and STO-RP$_2$, and 17 around 370 nm for STO-RP$_5$. Except for STO-RP$_1$, they are also very strong in the visible and near infrared spectral regions, approaching or exceeding 20 between 500 nm and 2 µm for STO-RP$_5$, STO-RP$_4$, STO-RP$_3$ and STO-RP$_2$. These values are comparable or even higher than the highest reported in the literature at any wavelength, namely 19 at 1900 nm for BaTiS$_3$.[34]

## 3. Conclusion

In summary, the refractive indices of STO-RP$_N$ thin layers grown on SrTiO$_3$ substrates are strongly anisotropic. These materials present birefringence and dichroism comparable to or higher than the highest values reported in the literature for bulk materials. In contrast to known highly anisotropic materials, STO-RP$_N$ optical anisotropy can be tuned by changing the RP order, which leverages flexibility to tune the material properties depending on the targeted functionality/device design. RP phase structuration combines the advantages of optical metamaterials in terms of permittivity engineering flexibility with the integrability of homogeneous anisotropic materials, thus bridging the gap between both approaches. In the end, STO-RP$_N$ thin layers are single crystalline and epitaxial, can be grown by using industry-standard growth processes. Besides, as STO-RP$_N$ thin layers can be epitaxially grown on Si and GaAs using SrTiO$_3$ templates,[76,77,78] our work paves the way for their compact integration on these photonic platforms. This collection of unique properties confers on this material considerable interest for the design of novel integrated applications for biomedical molecular recognition, optical communications, polarization detection or even quantum optics.

## 4. Experimental section:

*Sample growth:* The STO-RP$_N$ (N = 1,…5) thin layers were grown on (001)-oriented STO substrates in an oxide-MBE reactor. Ti and Sr were evaporated using Knudsen effusion cells and the growth was monitored using RHEED. Prior to growth, the substrates were annealed in the MBE reactor for 30 min at 700°C under an oxygen partial pressure of $10^{-7}$ Torr. The RP thin layers were then grown under the same pressure and temperature conditions. SrO and TiO$_2$ atomic layers were grown alternatively, layer by layer, at an approximative growth rate of 1 monolayer (ML) per minute. A growth sequence leveraging the swapping effect described in Ref.79 was used to optimize the thin layer structural quality.

*Structural characterizations:* X-ray reflectivity (XRR) and XRD experiments were performed using a Rigaku Smartlab diffractometer equipped with a high brilliance rotating anode and Ge(220)x2 monochromator placed on the incident beam. Details of the analysis of these measurements are given in the Supporting Information, section SII. The HAADF-STEM micrographs displayed in Fig.1 were recorded using a recorded using a JEOL JEM-ARM200F Cold FEG NeoARM 60-200kV probe corrected microscope, operated at 200kV and equipped with a spherical aberration corrector (Cs-probe CEOS ASCOR).

*Ellipsometry and reflectivity measurements:* The variable angle spectroscopic ellipsometry (VASE) measurements were carried out in the 260-2100 nm spectral range using a HORIBA scientific UVISEL plus ellipsometer. The ellipsometric parameters $\Delta$ and $\psi$ were recorded at 29 different incident angles ranging from 55 to 83°, except for the STO substrate for which the ellipsometric measurement was performed at a single incidence angle of 60°. The zero-incidence reflectivities $R_{0°}^{exp}$ were measured using an Ocean Optics NanoCalc reflectometer (spectral range 190 to 2100 nm).

## Supporting Information

Supporting Information is available from the Wiley Online Library or from the author.

## Acknowledgements

The authors acknowledge funding from the French Agence Nationale de la Recherche (ANR), CEAS-OFM project, grant number ANR-21-CE24-0003

**Conflict of Interest**

The authors declare no conflict of interest.

**Data Availability Statement**

The data that support the findings of this study are available from the corresponding author upon reasonable request.